\documentclass{PoS}
\pdfoutput = 1
\usepackage{cite}
\usepackage{amsmath}  
\usepackage{amssymb} 
\usepackage{dsfont}
 \usepackage{multirow}
\usepackage{psfrag,scalefnt}
\usepackage{paralist}

\title{Searching for New Physics through correlations of Flavour Observables}

\ShortTitle{Searching for NP through correlations of Flavour Observables}

\author{\speaker{Jennifer Girrbach}\\
        TU Munich, Institute for Advanced Study\\
        E-mail: \email{jennifer.girrbach@tum.de}}

\abstract{The coming flavour precision era will allow to uncover various patterns
of flavour violation in different  New Physics scenarios. We discuss different classes of them.
A simple extension of the Standard Model  that generally introduces new sources of flavour and CP violation as well as
right-handed currents is
the addition of a $U(1)$ gauge symmetry to the SM gauge group. In such $Z^\prime$ models correlations
between various flavour observables emerge that could test and distinguish different $Z^\prime$ scenarios.
A concrete
model with flavour violating $Z^\prime$ couplings is the 331 model based on the gauge group $SU(3)_C\times SU(3)_L\times U(1)_X$. We also
study tree-level FCNCs mediated by heavy neutral scalars and/or
 pseudo-scalars $H^0(A^0)$. Furthermore the implications of an additional  approximate
global $U(2)^3$ flavour symmetry is shortly discussed. Finally a model with vectorlike fermions and flavour violating $Z$
couplings is
presented. We identify a number of correlations between various observables that
differ from those known from constrained minimal flavour violating (CMFV) models and that could test and distinguish these different
scenarios.  }

\FullConference{14th International Conference on B-Physics at Hadron Machines,\\
		April 8-12, 2013\\
		Bologna, Italy }

\begin{document}

\section{Introduction}\label{sec:intro}

One highlight of LHCb so far was the measurement of $\overline{\mathcal{B}}(B_{s}\to\mu^+\mu^-)^\text{exp} =
(3.2^{+1.5}_{-1.2}) \times 10^{-9}$ \cite{Aaij:2012ac,LHCbBsmumu}\footnote{The ``bar'' notation means that $\Delta \Gamma_s$ 
effects are taken into account.}. This has to be compared to the SM prediction  $
 \mathcal{B}(B_{s}\to\mu^+\mu^-)^\text{SM}= (3.25\pm 0.17)\cdot 10^{-9}$ \cite{Buras:2012ru} and 
$\overline{\mathcal{B}}(B_{s}\to\mu^+\mu^-)^\text{SM}= (3.56\pm 0.18)\cdot 10^{-9}$ \cite{Buras:2013uqa} without and with  
including $\Delta \Gamma_s$ effects. So far everything is 
consistent with the SM and the room that is left for new physics (NP) gets smaller.

A slight tension in the flavour data concerns $|\varepsilon_K|$ and $S_{\psi K_S}$ which is
related to the so-called
$|V_{ub}|$-problem. Both $|\varepsilon_K|\propto \sin2\beta |V_{cb}|^4$ and $S_{\psi K_S}$ can be used to determine $\sin2\beta$.
The value for $\sin2\beta$ derived from the experimental value of $S_{\psi K_S}$
is
much smaller that the one derived from $|\varepsilon_K|$ \cite{Lunghi:2008aa,Buras:2008nn}.
 The ``true'' value of the angle $\beta$ of the unitarity triangle depends on the value of $|V_{ub}|$ and $\gamma$.  However there is a
tension between the exclusive
and
inclusive determinations of  $|V_{ub}|$ \cite{Nakamura:2010zzi}:
\begin{align}
 &|V_{ub}^\text{incl.}| = (4.27\pm 0.38)\cdot 10^{-3}\,,\qquad |V_{ub}^\text{excl.}| = (3.38\pm 0.36)\cdot
10^{-3}\,.
\end{align}
\begin{compactitem}
 \item Scenario 1 (S1): If one uses the
exclusive value of $|V_{ub}| $  to derive $\beta_\text{true}$ and then calculates $S_{\psi
K_S}^\text{SM}=\sin2\beta_\text{true}$ one
finds agreement with the data whereas $|\varepsilon_K|$ stays below the data.
\item  Scenario 2 (S2): Using the inclusive  $|V_{ub}| $ as input for 
$\beta_\text{true}$, $S_{\psi K_S}$ is above the measurements while $|\varepsilon_K|$ is in agreement with the
data.
\end{compactitem}
 However one has to keep in mind the error on  $|\varepsilon_K|$ coming dominantly from the 
error of $|V_{cb}|$ and the error of the QCD factor $\eta_1$~\cite{Brod:2011ty}\footnote{In \cite{Buras:2013raa} we propose a 
method how the uncertainty in $\eta_1$ could be reduced using the experimental value of $\Delta M_K$.}. Since $S_{\psi 
K_S}$ is a rather clean
observable in the SM one would need a new CP violating phase in S2 to get agreement with the SM. When studying NP models it is
interesting to see if this $|\varepsilon_K|-S_{\psi K_S}$ tension can be solved in this particular model and if yes, which 
scenario is
chosen by the model. CMFV chooses for example S1 because there are no new phases whereas the 331 model which I will discuss below
chooses S2 as effects in $\varepsilon_K$ are rather small and a new phase enters $B_d$-mixing.

\section{Phenomenology of $Z^\prime$, new (pseudo) scalar $A^0/H^0$ and vectorlike fermions}

What is the first new particle beyond the Higgs to be discovered at the LHC? Is it a new heavy gauge boson, a heavy (pseudo) scalar
or a heavy vectorlike fermion? If it is too heavy for a direct discovery then we can only see it in high precision 
flavour
experiments. In the following I will first discuss a concrete model where FCNC are induced by a $Z^\prime$ \cite{Buras:2012dp}.
Afterwards this approach is extended to more general $Z^\prime$ and also scalar scenarios \cite{Buras:2013rqa,Buras:2013uqa}.
At the end I present a model with vectorlike fermions \cite{Buras:2013td}. A summary of the implications of LHCb measurements on 
further models like Little Higgs, Randall Sundrum, SUSY GUTs can be found in \cite{Buras:2012ts}.

\subsection{Concrete model with $Z^\prime$ FCNC: 331 model}

The 331 model is based on the gauge group  $SU(3)_C\times SU(3)_L\times U(1)_X$. In the breaking $SU(3)_L\times U(1)_X\to
SU(2)_L\times U(1)_Y$ to the SM gauge group a new heavy neutral gauge boson $Z^\prime$ appears that mediates FCNC already at tree
level. A nice theoretical feature is that we have an
explanation of why there are $N =3$ generations. This follows from the requirement of
anomaly cancellation and asymptotic freedom of QCD. Anomaly cancellation is only possible if one generation (usually the
3$^\text{rd}$ is chosen) is treated differently than the other two generations. 

\subsubsection*{Flavour structure}

There are different versions of the 331 model characterized by a parameter $\beta$ that determines the particle content. We
consider $\beta = 1/\sqrt{3}$  (to be called $\overline{331}$ model) with the following fermion content: 
Left-handed fermions fit in (anti)triplets, while right-handed ones are singlets under $SU(3)_L$.
In the quark sector, the first two generations  fill the two upper components of a triplet, while the third one fills those
of an  anti-triplet; the  third member of the quark (anti)triplet is a new heavy fermion:
\begin{align}
&
 \begin{pmatrix}
                 e\\
-\nu_e\\
\nu_e^c
                \end{pmatrix}_L\,,
\begin{pmatrix}
                 \mu\\
-\nu_\mu\\
\nu_\mu^c
                \end{pmatrix}_L\,,
\begin{pmatrix}
                 \tau\\
-\nu_\tau\\
\nu_\tau^c
                \end{pmatrix}_L\,,\quad\qquad\begin{pmatrix}
           u\\d\\D
          \end{pmatrix}_L\,,
\begin{pmatrix}
 c\\s\\S
\end{pmatrix}\,,\begin{pmatrix}
       b\\-t\\T
      \end{pmatrix}_L\\
&
e_R, \,\mu_R,\, \tau_R,\,\qquad u_R,\, d_R,\, c_R,\, s_R,\, t_R,\, b_R,\,\qquad D_R, S_R, T_R
\end{align}
Due to anomaly cancellation we need the same number of triplets and anti-triplets. If one takes into account the three colours of
the quarks we have six triplets and six anti-triplets with this choice. 
Neutral currents mediated by  $Z^\prime$ are affected by the quark mixing since the $Z^\prime$ couplings are generation
non-universal.  In order to see this explicitly, we look at $Z^\prime$
couplings to SM quarks:
\begin{align}
 &\mathcal{L}^{Z^\prime}= J_\mu Z^{\prime\mu}\,,\quad\quad V_\text{CKM}=U_L^\dagger V_L,\\
&J_\mu = \bar u_L\gamma_\mu U_L^\dagger\begin{pmatrix}
                                        a& & \\ & a &\\ & & b
                                       \end{pmatrix}U_L u_L + \bar d_L\gamma_\mu V_L^\dagger\begin{pmatrix}
                                        a& & \\ & a &\\ & & b
                                       \end{pmatrix}V_L d_L\,.
\end{align}
The unitary rotation matrices $U_L, V_L$ do not cancel out for $a\neq b$ but generate tree-level FCNCs $\propto (b-a)$. However 
only left-handed (LH) quark
currents are flavour-violating. In Sec.~\ref{sec:general} we will generalize
this and include also right-handed (RH) currents. We choose a parametrization for $V_L$ using 3 angles and 3 phases $\tilde
s_{12}, \tilde s_{23}, \tilde s_{13}, \delta_{1,2,3}$  such that the $B_d$ sector depends only on $\tilde s_{13},\delta_1$, the
$B_s$ sector on $\tilde s_{23},\delta_2$ but then the
$K$ sector is correlated and depends on the same angles $\tilde s_{13}, \tilde s_{23}$ and the phase difference
$\delta_2-\delta_1$. In  more general models in Sec.~\ref{sec:general} the $K$ sector is then decoupled from $B_{d,s}$ sector.

\subsubsection*{Finding optimal oases in parameter space}

\begin{figure}[!tb]
\begin{center}
  \includegraphics[width=0.38\textwidth] {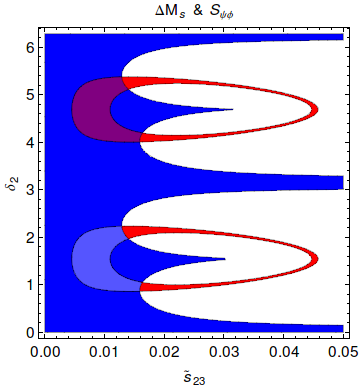}
\hspace{0.5cm}
   \includegraphics[width=0.38\textwidth] {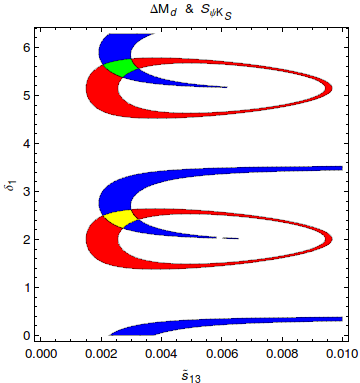}
\end{center}
\caption{Ranges for $\Delta M_{s,d}$ (red) and $S_{\psi \phi}/S_{\psi K_S}$ (blue) satisfying the bounds in
Eq.~(\protect\ref{equ:oases23}) and (\protect\ref{equ:oases13}).  The plot on the left is obtained for $M_{Z^\prime}=1$~TeV and
$|V_{ub} |=0.004$. The
purple, blue, green and yellow regions define our oases. }\label{fig:oases331}
\end{figure}

\begin{figure}[!tb]
 \centering
\includegraphics[width = 0.31\textwidth]{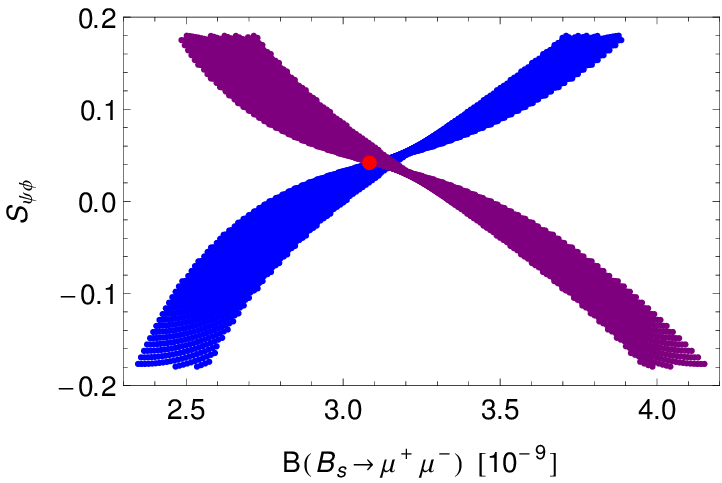}
\includegraphics[width = 0.31\textwidth]{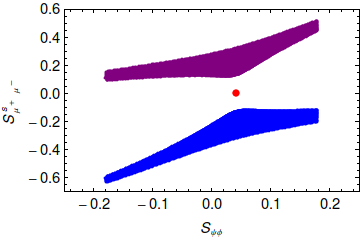}
\includegraphics[width = 0.31\textwidth]{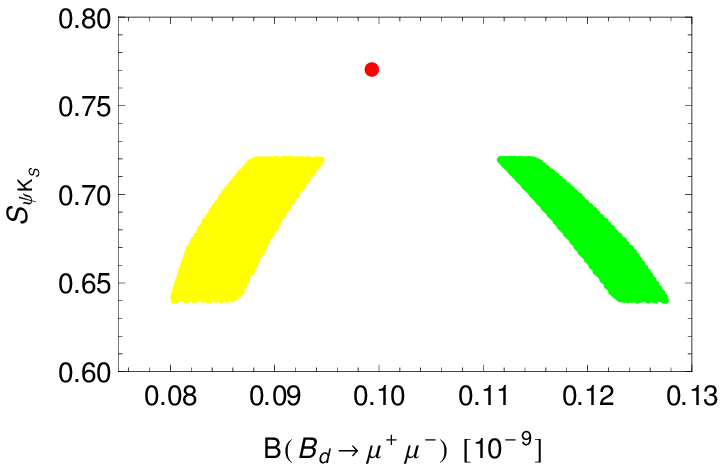}
\caption{$S_{\psi\phi}$ vs. $B_s\to\mu^+\mu^-$, $S_{\mu^+\mu^-}^s$ vs.  $S_{\psi\phi}$ and $S_{\psi K_S}$ vs. $B_d\to \mu^+\mu^-$
within the oases from Fig.~\protect\ref{fig:oases331} }\label{fig:DF1331}
\end{figure}

Our strategy that we will also adopt in the more general scenarios in Sec.~\ref{sec:general} is the following: we look first at $\Delta F =
2$ observables
to find constraints on the free parameters $\tilde s_{ij}$ and $\delta_i$. 
 In order to find these ``oases'' in the parameter space we require that the mixing induced CP asymmetries
are within their experimental 2$\sigma$ range and for the mass differences $\Delta M_{d,s}$ we take only 5\% error assuming a
flavour precision era ahead of us: 
\begin{align}\label{equ:oases23}
&16.9~{\rm ps}^{-1}\le \Delta M_s\le 18.7~{\rm ps}^{-1},
\quad  -0.18\le S_{\psi\phi}\le 0.18, \\
&0.48~{\rm ps}^{-1}\le \Delta M_d\le 0.53~{\rm ps}^{-1},\quad  0.64\le S_{\psi K_S}\le 0.72 . \label{equ:oases13}
\end{align}
Within these oases we then also include $\Delta F =1$
observables in order to find correlations between different observables. Such correlations can help to
identify and distinguish between different NP models.
In Fig.~\ref{fig:oases331} we show the two oases for $B_s$ and $B_d$ system, respectively. We use $M_{Z^\prime} = 1~$TeV and
$|V_{ub}| = 0.004$ such that $S_{\psi K_S}$ must be suppressed below its SM value. Due to the appearance of the phase $\delta_1$
this is indeed possible. The  two-fold ambiguity between $\delta_{1,2}$ and $\delta_{1,2} +\pi$ can be resolved with $\Delta F =
1$ observables (see Fig.~\ref{fig:DF1331}). Here $S_{\mu^+\mu^-}^s$ corresponds to a tagged time-dependent CP
asymmetry of $B_s\to\mu^+\mu^-$ \cite{DeBruyn:2012wk}.  Fig.~\ref{fig:DF1331} shows that we have a triple correlation
 $S_{\mu^+\mu^-}^s-S_{\psi\phi}-\mathcal{B}(B_s\to\mu^+\mu^-)$: once the sign of  $S_{\mu^+\mu^-}^s$ is known a unique
correlation $S_{\psi\phi}-\mathcal{B}(B_s\to\mu^+\mu^-)$ is found. If in
addition one of these three observables is precisely known the other
two can be strongly constrained. Effects in $B_d\to\mu^+\mu^-$ are rather small but one can see that in one oasis we find
enhancement and in the other a slight suppression w.r.t. SM central value. The new contributions in $K$ sector, especially in
$\varepsilon_K$, $K_L\to\pi^0\bar\nu\nu$ and $K^+\to\pi^+\bar\nu\nu$ turned out to be negligible. However the $S_{\psi
K_S}-\varepsilon_K$ tension explained in Sec.~\ref{sec:intro} could be solved using $|V_{ub}| = 0.004$.

\subsection{General $Z^\prime$ and $A^0/H^0$ scenarios}\label{sec:general}

\begin{figure}[!tb]
\centering
\includegraphics[width = 0.42\textwidth]{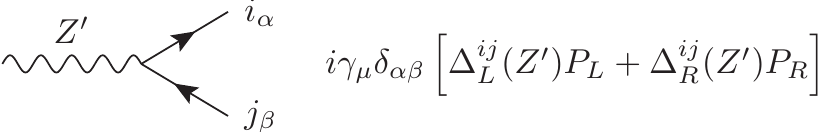}
\hspace{0.5cm}
\includegraphics[width = 0.42\textwidth]{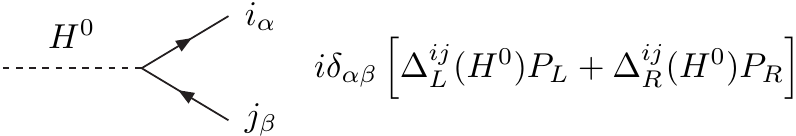}
\caption{Feynman rules for$Z^\prime$ and  neutral scalar particle $H^0$ ($i,\,j$ denote different
quark flavours and $\alpha,\,\beta$ the colours). }\label{fig:FR}
\end{figure}

The approach of the previous section is now generalized including both left-and right-handed $Z^\prime$/(pseudo)scalar FCNC
couplings (see Fig.~\ref{fig:FR}). We distinguish between different scenarios: LHS (RHS): $\Delta_{L(R)}\neq 0 = \Delta_{R(L)}$,
LRS: $\Delta_{L} = \Delta_{R}$ and ALRS $\Delta_{L} = -\Delta_{R}$. In addition we now also have to make assumptions about the
lepton couplings. In the 331 model this came out automatically from the Lagrangian:  $\Delta_L^{\nu\bar\nu}(Z')=0.14$ and
$\Delta_A^{\mu\bar\mu}(Z')=-0.26$. For the general $Z^\prime$ scenario
we set the lepton couplings at $\Delta_L^{\nu\bar\nu}(Z')=0.5$ and $\Delta_A^{\mu\bar\mu}(Z')=0.5$. 
In the SM both couplings of $Z$ are equal to $0.372$. In the scalar scenario we set $\Delta_P^{\mu\bar\mu}(H)= 0.020$ and
$\Delta_S^{\mu\bar\mu}(H)=0.040$ (for more details see \cite{Buras:2013rqa}).

Moreover we analyze what happens
if an additional $U(2)^3$ global flavour symmetry is imposed. In this case the $K$ system is governed by
MFV but $B_d$ and $B_s$ systems are now correlated \cite{Barbieri:2011ci}. Consequently instead of two separate oases plots for 
$B_d$ and $B_s$ system we
now have only one for all four observables. As pointed out in \cite{Buras:2012sd,Girrbach:2012gz} a triple correlation $S_{\psi
K_S}-S_{\psi\phi}-|V_{ub}|$ occurs such that now the $B_s$ oases depend also on $|V_{ub}|$.

\begin{figure}[!tb]
\centering
 \includegraphics[width= 0.38\textwidth]{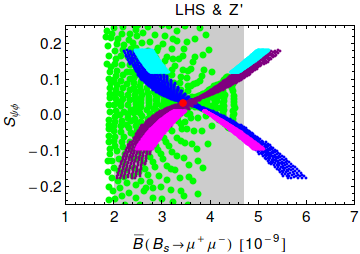}
\hspace{0.5cm}
 \includegraphics[width= 0.38\textwidth]{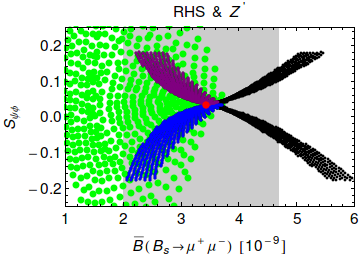}

 \includegraphics[width= 0.38\textwidth]{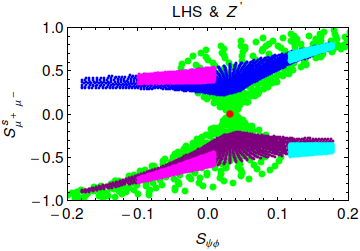}
\hspace{0.5cm}
\includegraphics[width = 0.36\textwidth]{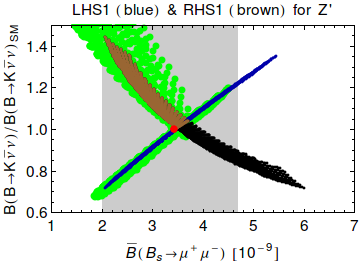}
\caption{$S_{\psi\phi}$ vs. $B_s\to\mu^+\mu^-$  in LHS (top left) and RHS (top right), 
$S_{\mu^+\mu^-}^s$ vs. $S_{\psi\phi}$ in LHS (down left) and $B\to K \nu\bar\nu$ vs $B_s\to\mu^+\mu^-$ for both LHS and RHS
(down right) for $M_{Z^\prime} = 1~$TeV. The green points indicate the regions that are compatible with $b\to s\ell^+\ell^-$ 
constraints of 
\cite{Altmannshofer:2012az}. Black points in RHS show the excluded area due to $b\to 
s\ell^+\ell^-$ transitions explicitly. Magenta and cyan
regions correspond to the $U(2)^3$ limit for small and large $|V_{ub}|$.}\label{fig:BsZprime1}
\end{figure}

In Fig.~\ref{fig:BsZprime1} the correlation $S_{\psi\phi}$ vs. $B_s\to\mu^+\mu^-$ for general $Z^\prime$ scenario is presented.
The difference between LHS and RHS is that the two oases are interchanged. Furthermore the black points in RHS show explicitly 
the regions that are excluded due to
$b\to s\ell^+\ell^-$ transitions where we used constraints derived in \cite{Altmannshofer:2012az}. The green points indicate the 
regions that are compatible with $b\to s\ell^+\ell^-$ 
transitions. As one can see these constraints restrict the RHS more than the LHS.  The magenta and cyan regions
corresponds to the $U(2)^3$ limit for $|V_{ub}| = 0.0031$ and $0.004$, respectively. For small $|V_{ub}|$ it follows that
$S_{\psi\phi}$ is  mainly negative. We also show $S_{\mu^+\mu^-}^s$ vs. $S_{\psi\phi}$ which is very similar to the
middle plot in Fig.~\ref{fig:DF1331} of the 331 model, because it's both for left-handed flavour changing $Z^\prime$ currents. A
possibility to distinguish between LHS and RHS is through $b\to s\bar\nu\nu$ transitions (see down right in 
Fig.~\ref{fig:BsZprime1}) where the brown/black points correspond to RHS and the blue ones to LHS. 

\begin{figure}[!tb]
\centering
 \includegraphics[width= 0.38\textwidth]{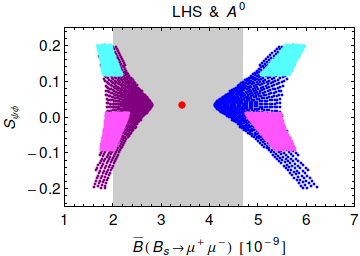}
\hspace{0.5cm}
 \includegraphics[width= 0.38\textwidth]{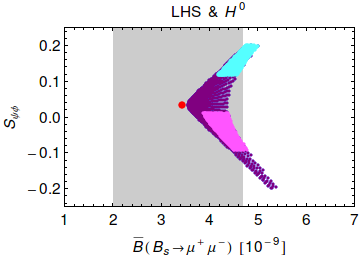}

 \includegraphics[width= 0.38\textwidth]{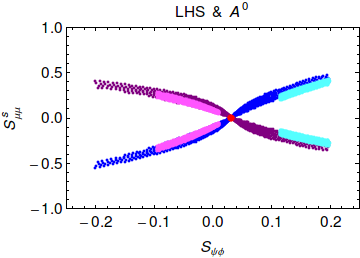}
\hspace{0.5cm}
 \includegraphics[width= 0.38\textwidth]{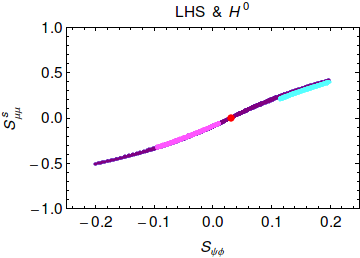}
\caption{$S_{\psi\phi}$ vs. $B_s\to\mu^+\mu^-$  in LHS (top) and 
$S_{\mu^+\mu^-}^s$ vs. $S_{\psi\phi}$ (down) for pseudoscalar $A^0$ (left) and scalar $H^0$ (right) in LHS
for $M_H = 1~$TeV.}\label{fig:Bsscalar}
\end{figure}

 \begin{figure}[!tb]
\begin{center}
\includegraphics[width=0.38\textwidth]{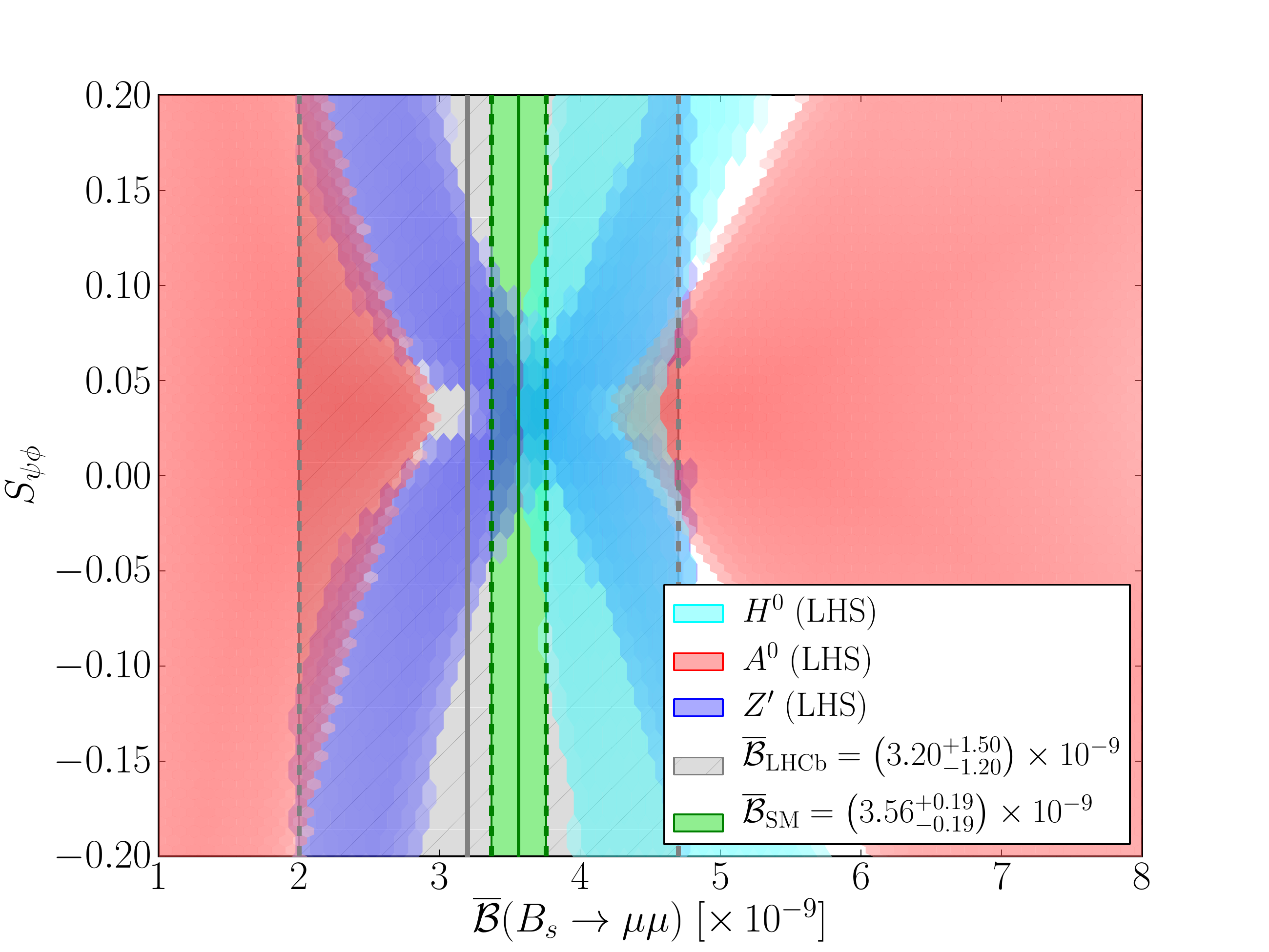}
\hspace{0.5cm}
\includegraphics[width=0.38\textwidth]{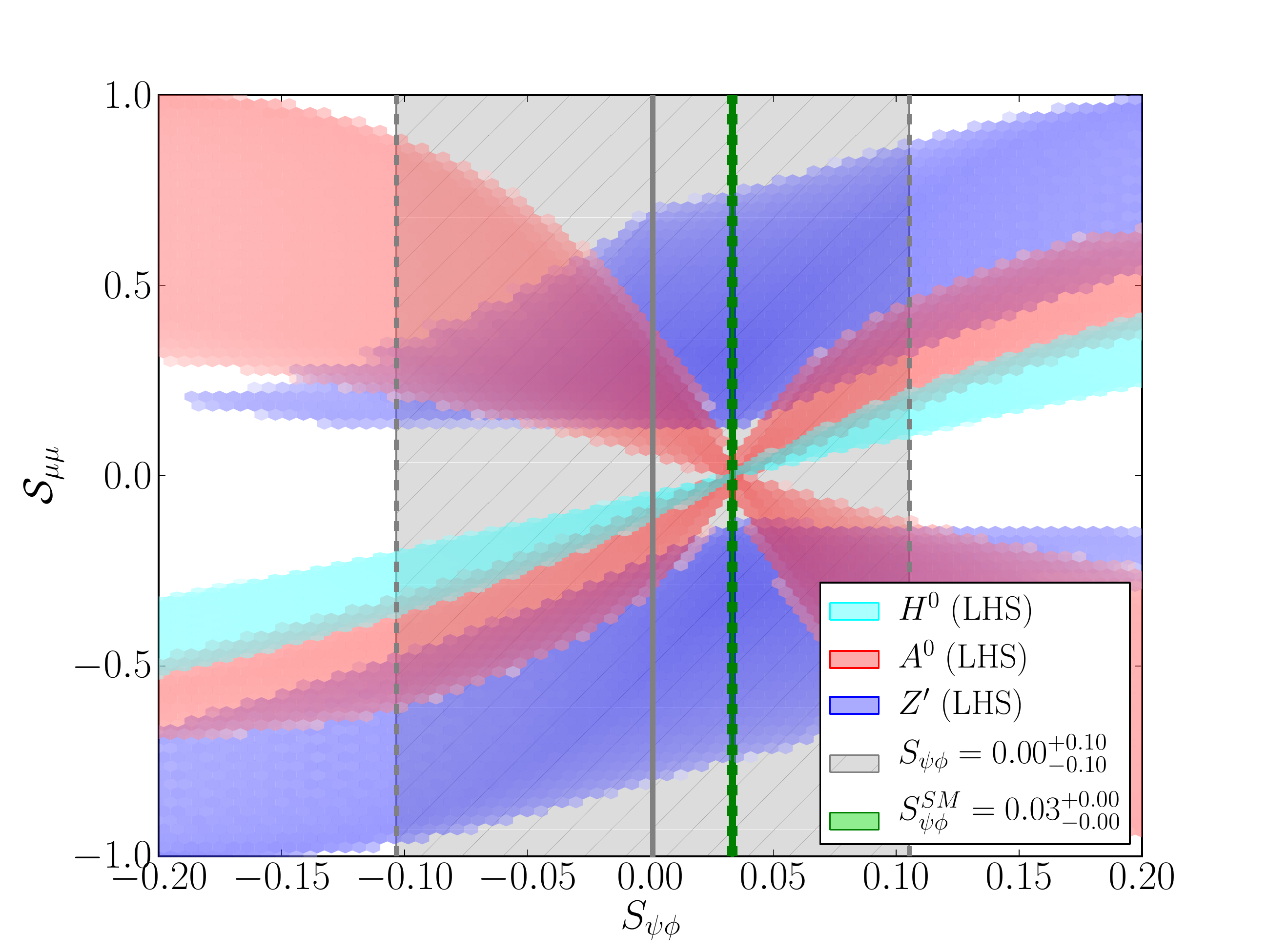}
\end{center}
\caption{Overlay of the correlations for $S_{\psi\phi}$ vs.
$B_s\to\mu^+\mu^-$ and $S_{\mu\mu}^s$ vs. $S_{\psi\phi}$   for tree
level scalar (cyan), pseudoscalar (red) and $Z^\prime$ (blue) exchange (both oases in same colour respectively) in LHS. The lepton
couplings are varied in the ranges $|\Delta_{S,P}^{\mu\mu}(H)| \in [0.02,0.04]$ and $\Delta_A^{\mu\mu}(Z')\in
[0.3,0.7]$.}\label{fig:grandplot}
\end{figure}

Correlations in the $B_s$ system for pseudoscalar and scalar scenario are shown in Fig.~\ref{fig:Bsscalar} where also the
$U(2)^3$ limit is included. In the scalar case the two oases cannot be distinguished and in $B_s\to\mu^+\mu^-$ only enhancement is
possible. In the pseudoscalar case both constructive and destructive interference with the SM contribution is possible and
effects can in principle be larger. The constraints from $b\to s\ell^+\ell^-$ transitions do not have any impact in the (pseudo) 
scalar case as shown in Ref.~\cite{Buras:2013rqa}. These results can now be compared with the corresponding plots for $Z^\prime$ 
case in 
Fig.~\ref{fig:BsZprime1}. We observe striking
differences between the results for  $Z^\prime$, $A^0$ and $H^0$ scenario due to their different spin and CP quantum numbers. A
summary plot is shown in Fig.~\ref{fig:grandplot} where now also the lepton couplings are varied in a wider range.

\begin{figure}
\begin{center}
\includegraphics[width=0.37\textwidth] {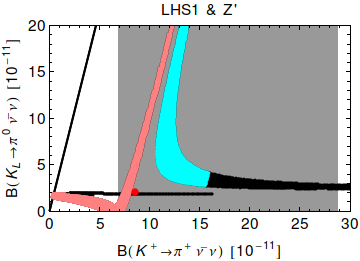}
\hspace{0.5cm}
\includegraphics[width=0.38\textwidth] {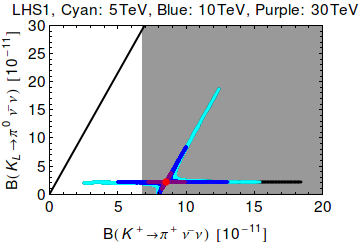}
\end{center}
\caption{$K_L\to\pi^0\nu\bar\nu$ vs. $K^+\to\pi^+\nu\bar\nu$ in LHS1 for $M_{Z^\prime} = 1~$TeV (left), 5/10/30~TeV (right). Black regions
are excluded by the upper bound $\mathcal{B}(K_L\to \mu^+\mu^-)\leq 2.5\cdot
10^{-9}$. Red point: SM central
value. Gray region:
experimental range of $\mathcal{B}(K^+\to\pi^+\nu\bar\nu)$.
}\label{fig:KZprime}
\end{figure}

Results for Kaon sector in $Z^\prime$ scenario are shown in Fig.~\ref{fig:KZprime}. Since only vector currents occur here there is no
difference between LHS and RHS. The deviations 
from the SM are significantly larger than in the case of
rare $B$ decays. This is a consequence of the weaker constraint from
$\Delta S=2$ processes compared to $\Delta B=2$ and the fact that rare $K$ decays
are stronger suppressed than rare $B$ decays within the SM. In $H^0/A^0$ scenario we expect negligible effects in channels with neutrinos.



\subsection{New vectorlike fermions: a minimal theory of fermion masses}\label{sec:vectorlike}


We now turn to a model with vectorlike fermions based on \cite{Buras:2011ph,Buras:2013td} that can be seen as a Minimal Theory of
Fermion Masses (MTFM). The idea is to explain SM fermion masses and mixings by their dynamical mixing with new heavy vectorlike
fermion $F$. Very simplified the Lagrangian has the following form: $\mathcal{L}\propto m \bar f F + M \bar F F + \lambda h F F$,
where $M$ denotes the heavy mass scale, $m$ characterizes the mixing and $\lambda$ is a Yukawa coupling. Thus the light fermions
have an admixture of heavy fermions with explicit mass terms. The Higgs couples only to vectorlike but not to chiral fermions, so
that SM Yukawas arise solely through mixing. We reduce the number of parameters such that
it is still possible to reproduce the SM Yukawa couplings and that at the same time
flavour violation is suppressed. In this way we can identify the minimal FCNC effects.
A central formula is the leading order expression for the SM quark masses
\begin{align}
m_{ij}^X = v \varepsilon_i^Q \varepsilon_j^X \lambda_{ij}^X\,,\qquad (X = U,D)\,,\qquad \varepsilon_i^{Q,U,D} =
\frac{m_i^{Q,U,D}}{M_i^{Q,U,D}} \,.
\end{align}
In \cite{Buras:2011ph} the heavy Yukawa couplings $\lambda^{U,D}$ have been assumed to be anarchical $\mathcal{O}(1)$ real 
numbers which allowed a first look at the phenomenological implications. In \cite{Buras:2013td} the so called TUM 
(Trivially Unitary Model) was studied in more detail. We assumed universality of heavy masses $M_i^Q = M_i^U = M_i^D = M$ and 
unitary Yukawa matrices. With this the flavour structure simplified considerably. Furthermore 
we concentrated on flavour violation in the down sector  and thus set $\lambda^U = \mathds{1}$. After fitting the SM quark masses 
and the CKM matrix  we are left with only four new real parameters and no new phases: $M,\,\varepsilon_3^Q,\, 
s_{13}^d,\, s_{23}^d$. The latter two parameters are angles of $\lambda^D$ (the third angle is fixed by the fitting procedure) 
and from fitting $m_t$ it follows that $0.8\leq \varepsilon_3^Q\leq 1$.

\begin{figure}[!tb]
\centering
\includegraphics[width = 0.38\textwidth]{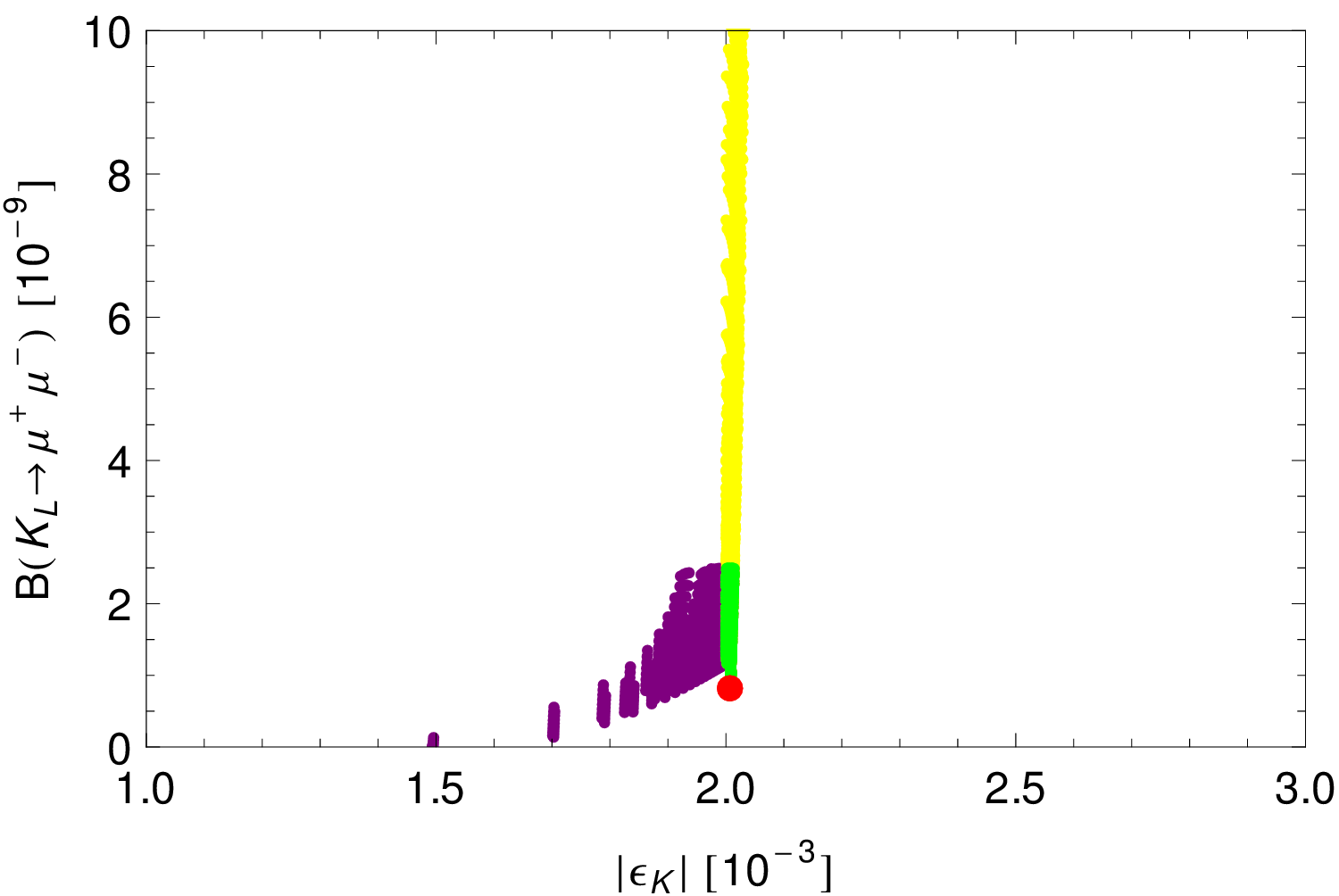}
\hspace{0.5cm}
\includegraphics[width = 0.38\textwidth]{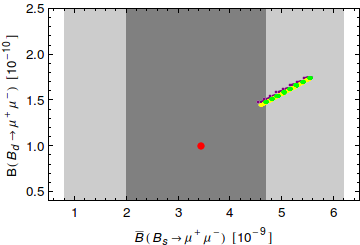}
\caption{$\mathcal{B}(K_L\to\mu^+\mu^-)$ vs. $|\varepsilon_K|$ and  $\mathcal{B}(B_d\to\mu^+\mu^-)$  vs. 
$\overline{\mathcal{B}}(B_s\to\mu^+\mu^-)$ for $M = 3~$TeV and $|V_{ub}| = 0.0037$. Green points are
compatible with both bounds for $|\varepsilon_K|$ (\protect\ref{C3}) and $\mathcal{B}(K_L\to\mu^+\mu^-)$ 
(\protect\ref{eq:KLmm-bound}), yellow is
only compatible with  $|\varepsilon_K|$ and purple only with $\mathcal{B}(K_L\to\mu^+\mu^-)$. The red
point corresponds to the SM central value. }\label{fig:TUM}
\end{figure}

The new contributions to FCNC processes are dominated
by tree-level flavour violating $Z$ couplings to quarks.  The simplest version of the MTFM, the TUM, is capable of describing the 
known quark mass spectrum and the elements of the CKM matrix favoring $|V_{ub} | \approx 0.0037$. Since there are no new phases 
$S_{\psi K_S}$ stays SM-like and thus the large inclusive value of $|V_{ub}|$ is disfavored. Although effects in $\varepsilon_K$ 
can in principle be large, the effects are bounded by 
\begin{equation}\label{eq:KLmm-bound}
\mathcal{B}(K_L\to\mu^+\mu^-)_{\rm SD} \le 2.5 \cdot 10^{-9}\,.
\end{equation}
For a $|V_{ub}|$ in between excl. and incl. value it is still possible to find regions in the parameter space that satisfy
\begin{equation}\label{C3}
0.75\le \frac{\Delta M_K}{(\Delta M_K)_{\rm SM}}\le 1.25,\qquad
2.0\times 10^{-3}\le |\varepsilon_K|\le 2.5 \times 10^{-3}
\end{equation}
and Eq.~(\ref{eq:KLmm-bound}) but then the prediction of the model is that $S_{\psi K_S}\approx 0.72$ which is by $2\sigma$ 
higher than its present experimental central value. In Fig.~\ref{fig:TUM} (left) we show the correlation 
$\mathcal{B}(K_L\to\mu^+\mu^-)$ vs. $|\varepsilon_K|$ for $M = 3~$TeV where only the green points satisfy (\ref{eq:KLmm-bound}) and 
(\ref{C3}) simultaneously. In the TUM effects in $B_{s,d}$ mixings are negligible  and the pattern of deviations from 
SM predictions in rare $B$ decays is 
CMFV-like as can be see 
on the right hand side of Fig.~\ref{fig:TUM}. However   
$\mathcal{B}(B_{s,d}\to\mu^+\mu^-)$ are uniquely enhanced  over their SM 
values. For $M=3~$TeV these enhancements amount to at least $35\%$ 
and can be as large as  a factor of two. With increasing $M$  the enhancements  decrease. However they remain 
sufficiently large for 
$M\le 5~$TeV to be detected in the flavour precision era.
Also effects in $K\to \pi\nu\bar\nu$ transitions are enhanced by a similar amount.
 
\section{Summary}

Correlations of flavour observables can help in identifying NP. We concentrated on rather simple extensions of the SM, namely 
those with tree-level FCNCs mediated by a $Z^\prime$, by an additional (pseudo) scalar $A^0/H^0$ or by the SM $Z^0$. The latter 
appeared in a model with new vectorlike fermions, the so-called TUM where $B_{d,s}\to\mu^+\mu^-$ are CMFV-like but enhanced. The 
$\overline{331}$ model is a concrete model with $Z^\prime$ FCNCs that are purely left-handed. Correlations of observables like 
$S_{\psi\phi}$, $B_s\to\mu^+\mu^-$ and $S_{\mu\mu}^s$ differ between $A^0$, $H^0$ and $Z^\prime$ case due to different spin and 
CP-parity which allow to distinguish between these scenarios. Rare Kaon decays $K\to\pi\nu\bar\nu$ play an important role even if the 
$Z^\prime$ mass is outside the reach of the LHC. One consequence of imposing an additional $U(2)^3$ flavour symmetry is that 
observables in the $B_s$ sector, especially the correlation between $S_{\psi\phi}$ and $B_s\to\mu^+\mu^-$ depends on $|V_{ub}|$
\cite{Buras:2012sd}.
With improved experimental data  and improved lattice
calculations these correlation will allow to
monitor how  the simple NP scenarios
face the future precision flavour data.

\acknowledgments

I am grateful for the invitation to Beauty 2013 and thank the organisers for the opportunity to give this talk. I thank all
my collaborators  A.~J.~Buras, M.~V.~Carlucci,  F.~De~Fazio,  R.~Fleischer, R.~Knegjens, M.~Nagai and R.~Ziegler for an enjoyable
collaboration. I thank A.~J.~Buras for proofreading this manuscript. The work presented in this talk is supported by the ERC Advanced 
Grant project ``FLAVOUR'' (267104) (ERC Report number 44).

\end{document}